\documentclass[10pt,conference]{IEEEtran}
\IEEEoverridecommandlockouts
\usepackage{cite}
\usepackage{amsmath,amssymb,amsfonts}
\usepackage{algorithmic}
\usepackage{graphicx}
\usepackage{textcomp}
\usepackage{xcolor}
\usepackage{comment}
\usepackage{fontawesome}
\usepackage{csquotes}
\usepackage{relsize,etoolbox}
\usepackage[rightmargin=0.5pt,leftmargin=0.5pt]{quoting}
\usepackage{booktabs} 

\usepackage{xspace}
\usepackage[normalem]{ulem}
\useunder{\uline}{\ul}{}
\usepackage{multirow}
\usepackage{lscape}
\usepackage{rotating}
\usepackage{supertabular}
\usepackage{longtable}
\usepackage{enumerate}
\usepackage{colortbl}
\usepackage{subcaption}
\usepackage{hhline}
\usepackage{makecell}
\usepackage{caption}
\usepackage{hyperref}
\usepackage{enumitem}
\usepackage{tcolorbox}
\usepackage{tabularx}
\usepackage{lipsum}
\usepackage{balance}
\usepackage{blindtext,graphicx}
\usepackage[absolute]{textpos}
\usepackage{fdsymbol}
\usepackage{url}

\usepackage{breakurl}
\usepackage{float}

\definecolor{my-color}{cmyk}{0.0, 0.0, 0.0, 0.0, 0.1}

\usepackage[T1]{fontenc}

\def\BibTeX{{\rm B\kern-.05em{\sc i\kern-.025em b}\kern-.08em
    T\kern-.1667em\lower.7ex\hbox{E}\kern-.125emX}}

\makeatletter 
\newcommand{\linebreakand}{%
  \end{@IEEEauthorhalign}
  \hfill\mbox{}\par
  \mbox{}\hfill\begin{@IEEEauthorhalign}
}
\makeatother 

\begin{document}

\title{A Study of Gender Discussions in Mobile Apps
}


\author{\IEEEauthorblockN{Mojtaba Shahin}
\IEEEauthorblockA{\textit{School of Computing Technologies} \\
\textit{RMIT University}\\
Melbourne, Australia \\
mojtaba.shahin@rmit.edu.au}
\and
\IEEEauthorblockN{Mansooreh Zahedi}
\IEEEauthorblockA{\textit{School of Computing and Information Systems} \\
\textit{University of Melbourne}\\
Melbourne, Australia \\
mansooreh.zahedi@unimelb.edu.au}
\linebreakand
\IEEEauthorblockN{Hourieh Khalajzadeh}
\IEEEauthorblockA{\textit{School of Information Technology} \\
\textit{Deakin University}\\
Melbourne, Australia \\
hkhalajzadeh@deakin.edu.au}
\and
\IEEEauthorblockN{Ali Rezaei Nasab}
\IEEEauthorblockA{\textit{School of Electrical and Computer Engineering} \\
\textit{Shiraz University}\\
Shiraz, Iran \\
alirezaei@hafez.shirazu.ac.ir}
}

\maketitle

\begin{abstract}
Mobile software apps (“apps”) are one of the prevailing digital technologies that our modern life heavily depends on. A key issue in the development of apps is how to design gender-inclusive apps. Apps that do not consider gender inclusion, diversity, and equality in their design can create barriers (e.g., excluding some of the users because of their gender) for their diverse users. While there have been some efforts to develop gender-inclusive apps, a lack of deep understanding regarding user perspectives on gender may prevent app developers and owners from identifying issues related to gender and proposing solutions for improvement. Users express many different opinions about apps in their reviews, from sharing their experiences, and reporting bugs, to requesting new features. In this study, we aim at unpacking gender discussions about apps from the user perspective by analysing app reviews. We first develop and evaluate several Machine Learning (ML) and Deep Learning (DL) classifiers that automatically detect gender reviews (i.e., reviews that contain discussions about gender). We apply our ML and DL classifiers on a manually constructed dataset of 1,440 app reviews from the Google App Store, composing 620 gender reviews and 820 non-gender reviews. Our best classifier achieves an F1-score of 90.77\%. Second, our qualitative analysis of a randomly selected 388 out of 620 gender reviews shows that gender discussions in app reviews revolve around six topics: \textit{App Features}, \textit{Appearance}, \textit{Content}, \textit{Company Policy and Censorship}, \textit{Advertisement}, and \textit{Community}. Finally, we provide some practical implications and recommendations for developing gender-inclusive apps.
\end{abstract}


\section{Introduction}

Our modern life and society more than ever depend on software systems. With 6.92 billion smartphone users worldwide (86.41\% of the current world population) \cite{smartphonenumber}, mobile software apps (apps) have the most diverse users among different types of software systems. The users of apps come from different ages, physical and cognitive capabilities, genders, educational levels, socio-economic backgrounds, religions, beliefs, etc \cite{khalajzadeh2022supporting, ramos2021considerations,fazzini2022characterizing}. Such characteristics are sometimes called human-centric issues or human-centric aspects in the software engineering community \cite{khalajzadeh2022supporting,fazzini2022characterizing}. Negligence of one or more of those human-centric aspects in any digital technologies, including apps, can lead to excluding some users and bringing difficulties to users and businesses (e.g., loss of reputation for businesses, reduced usage and acceptance of technology in society) \cite{hartzel2003self, fu2013people}. Hence, over the last few years, the goal of some governments, organisations, and practitioners has been to develop initiatives, policies, guidelines, and practices to design digital technologies working for all populations, which have been referred to by different names such as “inclusive design” and “universal usability” \cite{clarkson2013inclusive,stumpf2020gender}. 

Among these human-centric aspects that app developers and owners should consider while designing inclusive apps, gender is one of the most important ones \cite{stumpf2020gender,hartzel2003self,nunes2023gire,lopes2020gender}. Women comprise 49.7\% of the world population in 2022, while men are slightly more (50.3\%) \cite{worldpopulation}. Furthermore, the 2021 LGBT+ Pride Global Survey in 27 countries shows that 1\% of the surveyed adults defined themselves as “\textit{transgender, non-binary/non-conforming/gender-fluid or in another way rather than as male or female}” \cite{lgbtpopulation}. However, there are software systems, including apps, that have been accused of being gender biased, being unequal to a specific gender, or excluding some users based on their gender. For example, Buolamwini and Gebru \cite{buolamwini2018gender} found that three widely used commercial gender classification systems contain gender and skin bias, as they misclassify darker-skinned women more than lighter-skinned men. In two similar studies \cite{feine2020gender,west2019d}, it was shown that most text-based and voice-based conversational systems (e.g., Chatbots) are female by default (e.g., use female names) and prefer females over males. Several reasons can explain these gender-related issues in software systems, such as the lack of gender diversity in development teams, the inability or reluctance to collect and understand gender-related requirements and expectations, gender biases in the design and development processes, lack of gender diversity in user studies, etc. \cite{lopes2020gender,williams2014you,vasilescu2015gender, padala2020gender,nunes2023gire}.

There have been some works to investigate and improve gender equality and gender inclusiveness in software systems and/or reduce or avoid gender bias in software systems (e.g., \cite{nunes2023gire,vorvoreanu2019gender,noei2022study,padala2020gender,feine2020gender,terrell2017gender}). While these efforts are valuable, no systematic work focuses on understanding and classifying gender-related problems that an app might bring for its diverse users and gender-related requirements and features users might expect from apps. We argue that such an understanding can be the first and essential step to designing gender-inclusive apps and apps that support gender equality and diversity \cite{nunes2023gire}. App reviews (user reviews) include rich information that has been constantly mined and analysed by various researchers and practitioners to extract actionable information (e.g., bug reports, feature requests) for app developers and users \cite{maalej2015bug}. 

This work aims to uncover gender discussions in apps from the user perspective by analysing app reviews. Figure \ref{fig:examplereview} shows a review that the user expresses a gender-related concern. To achieve our goal, we first collected 7 million app reviews from 70 Android apps in Google Play Store and filtered out false positives with a keyword set. Next, we developed machine learning (ML) and deep learning (DL) classifiers with promising performance to distinguish gender reviews (i.e., reviews that include gender discussions) from non-gender reviews. Our best classifier (i.e., RoBERTa) achieves a precision of 86.64\%, recall of 95.31\%, F1-score of 90.77\%, accuracy of 90.31\%, and AUC (Area Under the Receiver Operating Characteristic Curve) of 94.93\%. We then qualitatively analysed the randomly chosen 388 out of 620 gender reviews. The analysis shows that gender discussions in app reviews revolve around six topics: \textit{App Features}, \textit{Appearance}, \textit{Content}, \textit{Company Policy and Censorship}, \textit{Advertisement}, and \textit{Community}.

\begin{figure}
    \centering
 \begin{tcolorbox}[sharp corners, boxrule=0.1mm, colback = my-color, left=1pt,right=1pt,top=1pt,bottom=1pt]
{\fontfamily{qcr}\selectfont
I will stop using this app because now some of the \textbf{emojis are only male}. Or perhaps supposed to be \textbf{gender neutral but they look like males}.}
\end{tcolorbox}
    \caption{An example of a gender review}
    \label{fig:examplereview}
\end{figure}

The key contributions of our paper are:
\begin{itemize}
    \item We are the first to investigate gender discussions in app reviews.
    \item We develop ML/DL classifiers to automatically detect gender reviews.
    \item We develop a deep understanding of the different types of gender discussions in app reviews.
    \item We construct a dataset of gender reviews and make it publicly available \cite{replication}.
    \item We report recommendations for the development and research of gender-inclusive apps.
\end{itemize}

In the rest of this paper, we summarise the related work in Section \ref{sec:background}. Section \ref{sec:reseachdesign} presents our research design. In Sections \ref{sec:automaticclassification} and \ref{sec:categoriegender}, we report the approaches and results of RQ1 and RQ2, respectively. We discuss and reflect on the main results in Section \ref{sec:discussion}, followed by outlining the possible threats and their corresponding solution in Section \ref{sec:Threats}. We conclude the paper in Section \ref{sec:conclusion}, along with some future research directions.

\section{Related Work}\label{sec:background}

In this section, we report on the literature investigating gender-related issues in (mobile) software development and the use of app reviews to understand gender biases.

\subsection{Gender Imbalance in (Mobile) Software Development}

Nunes et al. \cite{nunes2023gire} conducted a systematic mapping study to investigate how gender issues have been considered in software engineering. Results emphasise the impact of gender on development and products and reveal that there is a limitation of approaches for addressing gender imbalance in requirements engineering. The paper proposes a conceptual model to include and analyse gender-related concepts in the requirements elicitation and specification phases. Feine et al. \cite{feine2020gender} analysed gender bias in text-based conversational agents or chatbots design. The results reveal that most chatbots are designed in a gender-biased way as a female chatbot by their names and avatars, mostly in applications such as customer service and sales. 

Lopes et al. \cite{lopes2020gender} investigated how gender bias in mobile app development impacts user experience for different genders. The results show gender biases in design, which stem from a different definition of design priorities based on the participants’ and persona’s gender. The authors suggest the use of gender-neutral personas to identify balanced gender-related requirements in the design process and also develop tools to increase gender awareness in the design process. Padala et al. \cite{padala2020gender} investigated gender-biased barriers in open-source software projects and the impact of open-source software tools and infrastructure. The study found that the majority of the barriers the practitioners and newcomers identified contained some form of gender bias. At the same time, it was found that gender-biased barriers stated by women were notably more than the ones mentioned by men. The authors suggested changes to the open-source software environments and tools to alleviate this.

Several researchers have worked on detecting inclusivity issues in mobile apps. Burnett et al. \cite{burnett2016gendermag} developed GenderMag (Gender Inclusiveness Magnifier) to detect gender biases in software products. It works based on integrating a specialised cognitive walk-through with research-based personas to capture gender-related differences. 
InclusiveMag \cite{mendez2019gendermag} is an extension of GenderMag, which is a (meta-) method that enables generating systematic inclusiveness methods for different diversity dimensions. In \cite{mendez2019gendermag}, InclusiveMag is applied to a multi-case study that covers eight diversity dimensions of under-served populations in eight teams. While these works are merely detecting inclusivity and gender issues, Vorvoreanu et al. \cite{vorvoreanu2019gender} investigated the usefulness of GenderMag for designing less biased products. The researchers used GenderMag in an industrial software product and used the biases they found to derive design changes to improve the product. Empirical studies comparing the original and improved products decreased the gender gap and improved the success rate for women and men. 

A tool \cite{chatterjee2021aid} is developed to automate GenderMag in order to find gender-inclusivity related bugs in software products. The evaluation of the tool on 20 GitHub projects revealed extra inclusivity bugs that manual GenderMag was not able to detect. Guizani et al. \cite{guizani2020gender} and Hilderbrand et al. \cite{hilderbrand2020engineering} studied 10 software teams who used GenderMag to investigate the impact of gender inclusivity for creating gender inclusive software. Their results show that all the teams decided to fix their inclusivity bugs with eight having the bugs fixed, one having its follow-ups rejected, and one having the fixes in progress. The results also revealed a positive impact on the team member’s mindset and awareness. Kanij et al. \cite{kanij2022new} built on top of GenderMag and presented a method to find gender bias in software engineering job advertisements aiming to overcome gender bias in software development. The authors collected responses from 44 software practitioners and identified 16 differing factors among male and female participants. They derived three software engineering job applicant personas. A pilot study using the personas on four job advertisements revealed the gender biases in the job advertisements.

 Ramos et al. \cite{ramos2021considerations} presented a scoping review to assess the consideration of diversity, equity, and inclusion in mobile applications designed for mental health management, by the existing assessment tools. The study identified 44 unique app evaluation frameworks being used in 68 studies to evaluate a mental health app. Results show limited consideration for diversity, equity, and inclusion variables that can limit the applicability of those app-based interventions for marginalised communities. The paper provides recommendations to improve the current mobile app evaluation frameworks’ effectiveness on diversity, equity, and inclusion.

\subsection{Gender Bias in App Reviews}

Fazzini et al. \cite{fazzini2022characterizing} did an empirical study on 2,611  COVID-19 app reviews to understand the inclusivity issues reported by the users. They categorised them into nine categories of Age, Disability, Emotion, Gender, Language, Location, Privacy, Socioeconomic, and Miscellaneous. The findings of this paper show that gender-related issues are very rarely discussed in COVID-19 apps compared to other issues, e.g. privacy and language. Noei et al. \cite{noei2022study} investigated  Google Play Store user reviews to compare the participation of men and women in posting the reviews. Analysing 438,707 user reviews on 156 Android apps revealed that women rate apps more positively than men, and women’s reviews receive a higher ranking. However, the results show women leave fewer reviews, which makes their reviews less visible to app developers and other users. This leads to developers being less likely to respond and address the reviews left by women, and therefore change apps mostly based on the needs of men users. This further discourages the participation of women in reviewing apps on Google Play Store. The authors suggest taking gender into account when responding to reviews and further investigating the motivations of app reviewers and how developers respond to them.

Even though these papers have investigated gender bias in software development and gender differences in app reviews, no research has been conducted to automatically detect mobile app reviews that contain gender discussions and understand the types of gender discussions. In this paper, we have studied gender discussions from the user perspective by analysing app reviews.

\section{Research Design}\label{sec:reseachdesign}
This study aims to develop a deep understanding of discussions around gender in mobile apps. To realise this goal, we purpose to answer the following research questions (RQs):


\begin{tcolorbox}[sharp corners, boxrule=0.1mm,]
\textbf{RQ1}: Can we accurately and automatically detect reviews that include discussions about gender?
\end{tcolorbox}

\textbf{Motivation}. Apps, in particular popular apps such as social networking and gaming apps, can receive millions of reviews from their diverse users. Users talk about multiple topics in app reviews. Hence, manually analysing such large reviews to identify reviews that contain the topic of interest might be time-consuming and error-prone. While there are some automated classifiers for mining app reviews (e.g., \cite{9793551,tushev2022domain,chen2014ar,alshangiti2022hierarchical}), gender concerns have not been the focus of such automated classifiers. Furthermore, applying general-purpose automated classifiers may misidentify gender reviews. In this RQ, we develop and experiment with multiple ML/DL models to help app developers and owners automatically and accurately detect gender reviews that include gender discussions.


\begin{tcolorbox}[sharp corners, boxrule=0.1mm,]
\textbf{RQ2}: What types of discussions about gender are expressed in app reviews?
\end{tcolorbox}

\textbf{Motivation}. While the ML/DL models in \textbf{RQ1} help app developers distinguish gender reviews from non-gender reviews, they still need to understand the different gender-related concerns users describe. Such knowledge enables them to make data-driven decisions in the development of gender-inclusive apps. Hence, this RQ aims to understand and categorise the types of gender discussions in reviews.

\subsection{A Dataset of Gender Reviews}
\label{sec:dataset}
To answer RQ1 and RQ2, we first construct a dataset of gender reviews (reviews that include gender discussions). We followed three steps for this purpose (See Figure \ref{fig:reseachmethod}):
\begin{figure*}
    \centering
    \includegraphics[width=0.95\linewidth]{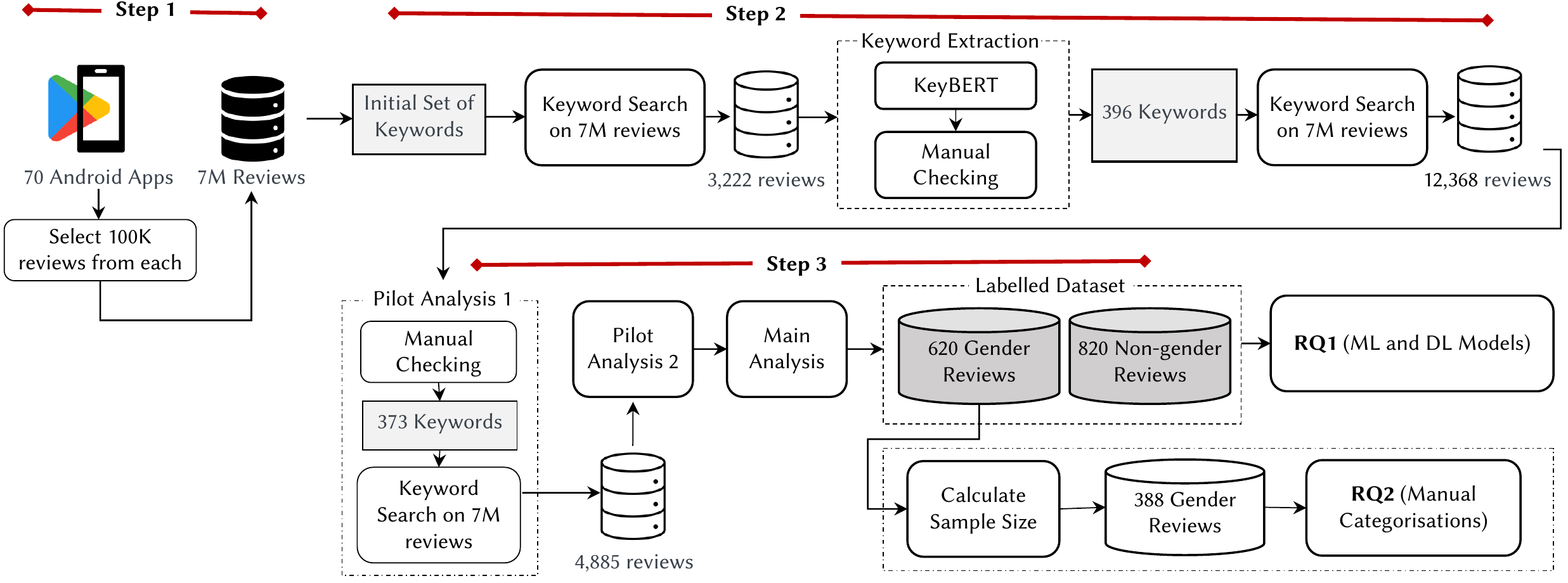}
    \caption{The process of constructing a dataset for RQ1 and RQ2}
    \label{fig:reseachmethod}
\end{figure*}

\textbf{Step 1. Collecting reviews of 70 apps}. We gathered the top 70 popular Android apps in the Google Play Store using App Annie Report \cite{annie2020app}. These 70 apps come from 22 diverse categories, such as communication, entertainment, social, finance, etc. Our replication package \cite{replication} provides information (e.g., the number of downloads and categories) on these apps. Google Play Store can display reviews given to an app in different ways, such as in chronological order or according to relevance. To avoid short reviews (e.g., one or two words long) or not very informative reviews, we relied on the most relevant reviews sorted by Google Play Store. As shown in Figure \ref{fig:reseachmethod}, for each app, we collected the top 100K most relevant app reviews (in total 7 million reviews) using a Google Play Crawler \cite{googleplaycrawler}.

\textbf{Step 2. Identifying potential gender reviews}. The 7 million reviews collected from 70 apps might discuss different topics (e.g., privacy). Hence, the next step was to identify reviews that were highly likely to include discussions around gender (we refer to them as \textit{\textbf{potential gender reviews}}). We began by constructing a relatively comprehensive set of keywords through a combination of manual and automated iterations, as described below. Our aim was twofold: (1) to minimise the number of uncontrolled false positives and (2) to ensure that relevant keywords were not excluded. 

\textit{\underline{Step 2.1 Creating an initial set of keywords}}. Consulting with (grey) literature (e.g., \cite{genderterms, feministterms,lgbtpopulation, hyde2019future}), we created an initial list of gender keywords, which include 45 uni-gram or two-gram keywords, such as “gender”, “gender inequality”, “feminism”, and “non-binary”. We then filtered our 7M reviews based on the 45 keywords and got 3,222 reviews. Note that we refrained to include some general gender terms such as “girl”, “men”, and “women” in this step as they could lead to a lot of false positives.

\textit{\underline{Step 2.2 Keywords extraction using KeyBERT}}. The initial set of keywords in the previous step might not be complete and comprehensive. Hence, as shown in Figure \ref{fig:reseachmethod}, we applied the KeyBERT technique \cite{grootendorst2020keybert} to extract more gender keywords.  KeyBERT is a BERT-based technique that leverages pre-trained BERT embedding models to extract the most similar n-gram keywords and phrases in a document \cite{keyBertsite}. Several pre-trained embedding models can be utilised in KeyBERT. We used the “all-mpnet-base-v2” embedding model as it is suitable for English documents \cite{pretrainedmodels}. We first performed KeyBERT on 3,222 reviews with a list of seed keywords as input to extract uni-gram, two-gram, and three-gram keywords from 3,222 reviews. The list of seed keywords was used to guide KeyBERT and includes the initial set of keywords (45 keywords) and 8 general gender-related keywords (e.g., “boy”, “men”, “female”, and “women”). Similarly, we then separately applied KeyBERT on three sexist datasets that were built based on Twitter data \cite{rodriguez2021overview,
samory2021call,
jha2017does}. Next, we manually checked the keywords suggested by KeyBERT (i.e., KeyBERT was performed on four datasets). Here, we mainly removed duplicates and deleted the extracted 2-gram or 3-gram keywords (e.g., “gender imbalance”) that we had their uni-gram keywords (e.g., “gender”). Note that in this step, very few extracted keywords (less than 2\%) were removed because we thought they were irrelevant. This process led to the identification of 396 gender keywords. We refer to these 396 gender keywords as our keyword set. Next, we filtered our initial 7M reviews using the keyword set and got 12,368 reviews, which we call \textbf{\textit{potential gender reviews}}. The rest of the reviews (6,987,632 = 7M - 12,368 ) that did not contain any of our gender keywords are called \textbf{\textit{non-gender reviews}}.

\textbf{Step 3. Constructing a dataset of gender reviews}. Despite applying the above-mentioned approaches, potential gender reviews (12,368 reviews) might include many false positives. Hence, we aimed to exclude false positives as much as possible from potential gender reviews and build a dataset of \textit{\textbf{gender reviews}} (reviews that include discussions around gender). We performed the following three steps for this purpose (See Figure \ref{fig:reseachmethod}):

\textit{\underline{Step 3.1 Pilot analysis 1}}. We randomly selected 100 reviews from 12,368 potential gender reviews and 100 reviews from 6,987,632 (7M - 12,368) non-gender reviews. Three analysts (three authors) with over five years of experience in research on human and social aspects of software participated in the analysis. They were asked to independently read these 200 reviews and classify the reviews as gender reviews or non-gender reviews. Then, the analysts held a Zoom meeting to discuss the results and disagreements. In total, 29 disagreements were found among the three analysts, which were settled through the negotiated agreement approach \cite{campbell2013coding}. This process led to identifying 22 gender reviews and 178 non-gender reviews, indicating that only 22 out of the 100 chosen potential gender reviews were gender reviews. Our main observation from this pilot analysis was that some of the uni-gram keywords in our keyword set can cause a lot of false positives. To solve this problem, the first author manually checked 10 reviews corresponding to every uni-gram keyword. In case a uni-gram keyword resulted in less than 10 reviews, the first author checked all of them. This led to removing 23 uni-gram keywords (e.g., “discrimination”, “prejudice”) from the keyword set (396 - 23). It is because none of the selected reviews corresponding to these 23 uni-gram keywords was about gender at all or if they were related to gender, they included other keywords in our keyword set. Hence, we argue that removing these 23 uni-gram keywords cannot seriously threaten the final dataset's representativeness. We next searched the keywords (373) in the refined keyword set on 7M reviews and got 4,885 potential gender reviews.

\textit{\underline{Step 3.2 Pilot analysis 2}}. In this step, we randomly chose 150 reviews from 4,885 potential gender reviews and 50 reviews from 6,995,150 (7M - 4,885) non-gender reviews. Similar to pilot analysis 1, the three analysts manually classified the 200 reviews as gender reviews or non-gender reviews. After resolving 43 disagreements, 92 gender reviews and 108 non-gender reviews were found, indicating that 92 out of the 150 chosen potential gender reviews (61.3\%) were gender reviews. This gave us enough confidence that the remaining 4,685 potential gender reviews (4,885 - 200) include a reasonable number of gender reviews.


\textit{\underline{Step 3.3 Main analysis}}. At the end of the pilot analysis, we had 114 gender reviews, 286 non-gender reviews, and 4,685 potential gender reviews. The 4,685 potential gender reviews belonged to 55 apps, of which 52 apps had at least 20 reviews. To collect balanced data from the highest number of possible apps and not be biased toward some specific apps, we randomly selected 20 reviews from each of these 52 apps that had at least 20 reviews (1,040 reviews). Like the pilot analysis, the analysts reviewed 1,040 reviews in three iterations independently (each iteration 347 reviews) and categorised them as 506 gender reviews and 534 non-gender reviews. 126 disagreements were found in this step and were settled using the negotiated agreement approach. At the end of the pilot and main analysis processes, we had a dataset of 620 gender reviews (506+114) and 820 (534+286) non-gender reviews.

\section {Automatic Detection of Gender Reviews (RQ1)} \label{sec:automaticclassification}
\subsection{Approach} \label{sec:approachautomatic}
This section presents 6 ML and 3 DL models that automatically identify gender reviews. 
\subsubsection{Dataset} We used the dataset created in Section \ref{sec:dataset}, consisting of 620 gender reviews and 820 non-gender reviews to evaluate ML and DL classifiers. We performed a very basic pre-processing on the dataset. We removed emojis, numbers, punctuations, and one-long characters.

\subsubsection{Feature Selection} We used three widely used text feature extraction methods including TFIDF (“Term Frequency–Inverse Document Frequency”), USE (“Universal Sentence Encoder”), and W2V (Word2Vec) to extract features from reviews in our dataset. These feature extraction techniques and their combination are used as input for ML and DL models and are briefed below.

\textbf{TFIDF} is an extensively used method in natural language processing (NLP), which computes the weight of a word (i.e., the importance of a word) in a collection of text documents (corpus) by multiplying the frequency of the emergence of the word in the document (TF) and the inverse document frequency (IDF) of the word across the corpus (TF*IDF) \cite{ramos2003using}. 
In our TFIDF, we use a feature union of TFIDF weighted with character 5-gram and top 50,000 features \cite{vogel2020fake}.
    
\textbf{W2V} is a word embedding technique that uses a neural network model to process a corpus by converting its words into a word vector \cite{mikolov2013efficient}. The resulting numerical word vectors can be used as features in ML and DL models. We leveraged the 300-dimensional Google News word2vec embedding model \cite{mikolov2013distributed} to generate a vector for each word in our dataset consisting of gender and non-gender reviews.    

\textbf{USE}. In contrast to word embedding techniques that use pre-trained \textit{word} embedding models, USE as a deep neural network-based model works at the \textit{sentence level} and leverages pre-trained \textit{sentence} embedding models to generate sentence vectors \cite{cer2018universal}. Transformer \cite{vaswani2017attention} and Deep-Averaging-Network (DAN) \cite{iyyer2015deep} are two available encoding models in USE that can be used to encode sentences into vectors. Note that USE in this study encodes a sentence to a 512-dimensional vector using the DAN encoder.

\textbf{Combinations}. We also used the various combinations of the above-mentioned feature techniques to explore if such combinations ultimately lead to improving the performance of ML and DL classifiers (See Table \ref{tbl:MLResults}).

\subsubsection{Classification Models} \label{sec:models} To classify app reviews into gender reviews and non-gender reviews, we used both ML and DL models. Our decision to use different classifiers was motivated by the following points: (1) it is challenging to select a classifier that yields the optimal results from the beginning. (2) We wanted to use both traditional ML classifiers and the state-of-the-art DL classifiers that are extensively used and have shown promising performance in text classification tasks \cite{aggarwal2012survey}. (3) ML/DL classifiers may yield different results as they make different assumptions on the examined dataset \cite{caruana2006empirical,abdalkareem2020machine}. (4) ML/DL classifiers have different execution times and strategies for dealing with overfitting and explainability \cite{kotsiantis2006machine, abdalkareem2020machine}.
We used 6 well-known and widely used ML models in software engineering: Logistic Regression \cite{dreiseitl2002logistic}, SVM-LR (Support Vector Machine based on Linear Kernel) \cite{cortes1995support}, Random Forest \cite{breiman2001random}, XGBoost (Extreme Gradient Boosting) \cite{chen2016xgboost}, AdaBoost (Adaptive Boosting) \cite{freund1997decision}, Decision Tree \cite{quinlan1986induction}. Our replication package \cite{replication} describes how we fine-tuned the parameters of these ML classifiers.


The DL models below are used to identify gender reviews and all of them are optimised using the Adam algorithm \cite{kingma2014adam}. 

\textbf{CNN} (Convolutional Neural Network) typically includes several convolution layers, one max pooling layer, and one or two fully connected layers \cite{yamashita2018convolutional}. Our designed CNN model includes one convolution layer with a feature map of 256, a kernel size of 5-gram, and the activation of ReLU (Rectified Linear Unit) \cite{ide2017improvement}. The CNN model also includes one max pooling layer with the size of 2 and one fully connected layer. The convolution and pooling layers are responsible to extract features from reviews and the fully connected layer categorises a review into a gender review or non-gender review. 

\textbf{LSTM} (Long Short-Term Memory) is a class of recurrent neural systems and has shown promising results in different domains such as NLP and time series processing. In contrast to standard recurrent neural networks (RNNs), LSTM is designed to address long-term dependencies in the data (e.g., images, text) \cite{hochreiter1997long, van2020review}. Our LSTM model is structured into one LSTM layer with a dimensionality factor of 256 and one max pooling layer with a size of 2.   

\textbf{RoBERTa} (Robustly optimised BERT approach) is a state-of-the-art classifier that extends BERT (Bidirectional Encoder Representations from Transformers \cite{devlin2018bert}) \cite{liu2019roberta}. RoBERTa can be applied to a variety of NLP tasks and outperforms BERT by modifying and extending some of BERT's design decisions such as training with larger data and with longer sequences. 

\subsubsection{Performance Measures}
We used five well-known and widely used metrics including precision, recall, F1-score, accuracy, and AUC to assess the performance of the ML and DL models \cite{yang2022survey, abdalkareem2020machine,viviani2019locating}. These metrics work based on four factors: False Positive (FP), True Positive (TP), True Negative (TN), and False Negative (FN). 
In our study, the precision metric determines the percentage of identified gender reviews that are actually gender reviews (i.e., $Precision= \frac{TP}{TP+FP}$). The recall metric is the ratio between the number of precisely labelled gender reviews and the total number of gender reviews (i.e., $Recall= \frac{TP}{TP+FN}$). We calculate the value of the F1-score metric by combing precision and recall (i.e., $F1-score = \frac{2*Precision*Recall}{Precision + Recall}$). Accuracy is computed as the ratio of reviews correctly labelled as gender reviews and non-gender reviews to the total number of reviews (i.e., $Accuracy= \frac{TP+TN}{TP+FP+TN+FN}$). Finally, AUC computes the likelihood that a classifier will classify a randomly selected gender review (i.e., TP) higher than a randomly selected non-gender review (i.e., FP). 

The 10-fold cross-validation method was employed to appraise the ML and DL classifiers \cite{efron1983estimating}. The 10-fold cross-validation method randomly divides our dataset, including 620 gender reviews and 820 non-gender reviews, into ten separate folds and processes the dataset 10 times. In each iteration, one fold of data (10\%) is picked up for assessment, and the remaining data (i.e., 90\%) is considered to train the classifier.

\subsection{Results}\label{sec:resultsautomatic}

\begin{table}[]
\caption{Results of ML/DL models (in \%): The best results of each metric are grayed and bold. Feat.: Features, PR: Precision, RE: Recall, FS: F1-score, ACC: Accuracy.}
\label{tbl:MLResults}
\setlength{\tabcolsep}{5pt}
\renewcommand{\arraystretch}{1.1}
\centering
\begin{tabular}{|ll|c|c|c|c|c|}
\hline
\multicolumn{1}{|l|}{\textbf{Feat.}}               & \textbf{Models} & \textbf{PR}                   & \textbf{RE}                   & \textbf{FS}                   & \textbf{ACC}                  & \textbf{AUC}                  \\ \hline
\multicolumn{1}{|l|}{}                                & LSTM            & 60.92                         & 69.55                         & 64.41                         & 67.84                         & 70.54                         \\ \cline{2-7} 
\multicolumn{1}{|l|}{}                                & CNN             & 74.67                         & 79.62                         & 76.86                         & 76.57                         & 84.64                         \\ \cline{2-7} 
\multicolumn{1}{|l|}{}                                & SVM-linear      & 81.74                         & 82.19                         & 81.82                         & 82.06                         & 89.38                         \\ \cline{2-7} 
\multicolumn{1}{|l|}{}                                & Random Forest   & 81.51                         & 85.65                         & 83.34                         & 83.12                         & 90.31                         \\ \cline{2-7} 
\multicolumn{1}{|l|}{\multirow{-5}{*}{\rotatebox[origin=c]{90}{\parbox[c]{0.8cm}{TFIDF}}}}         & XGBoost         & 83.66                         & 84.8                          & 84.13                         & 84.01                         & 90.76                         \\ \hline
\multicolumn{1}{|l|}{}                                & LSTM            & 67.46                         & 97.26                         & 79.33                         & 74.97                         & 90.36                         \\ \cline{2-7} 
\multicolumn{1}{|l|}{}                                & CNN             & 84.89                         & 87.46                         & 85.87                         & 85.62                         & 92.77                         \\ \cline{2-7} 
\multicolumn{1}{|l|}{}                                & SVM-linear      & 84.47                         & 87.86                         & 85.91                         & 85.78                         & 93.22                         \\ \cline{2-7} 
\multicolumn{1}{|l|}{}                                & Random Forest   & 85.79                         & 83.57                         & 84.58                         & 85.05                         & 93.06                         \\ \cline{2-7} 
\multicolumn{1}{|l|}{\multirow{-5}{*}{\rotatebox[origin=c]{90}{\parbox[c]{0.5cm}{USE}}}}           & XGBoost         & 84.79                         & 87.34                         & 85.98                         & 86.02                         & 93.56                         \\ \hline
\multicolumn{1}{|l|}{}                                & LSTM            & 63.61                         & 93.89                         & 75.23                         & 69.22                         & 84.07                         \\ \cline{2-7} 
\multicolumn{1}{|l|}{}                                & CNN             & 79.66                         & 81.86                         & 80.24                         & 80.69                         & 88.6                          \\ \cline{2-7} 
\multicolumn{1}{|l|}{}                                & SVM-linear      & 79.2                          & 77.46                         & 78.14                         & 78.67                         & 86.1                          \\ \cline{2-7} 
\multicolumn{1}{|l|}{}                                & Random Forest   & 78.5                          & 77.36                         & 77.67                         & 78.18                         & 86.28                         \\ \cline{2-7} 
\multicolumn{1}{|l|}{\multirow{-5}{*}{\rotatebox[origin=c]{90}{\parbox[c]{0.5cm}{W2V}}}}           & XGBoost         & 78.13                         & 79.49                         & 78.68                         & 78.91                         & 87.96                         \\ \hline
\multicolumn{1}{|l|}{}                                & LSTM            & 71.16                         & 93.93                         & 80.23                         & 76.57                         & 82.25                         \\ \cline{2-7} 
\multicolumn{1}{|l|}{}                                & CNN             & 85.41                         & 86.08                         & 85.56                         & 85.86                         & 92.54                         \\ \cline{2-7} 
\multicolumn{1}{|l|}{}                                & SVM-linear      & 85.19                         & 89.53                         & 87.23                         & 87.07                         & 94.24                         \\ \cline{2-7} 
\multicolumn{1}{|l|}{}                                & Random Forest   & 85.63                         & 86.01                         & 85.65                         & 85.86                         & 93.04                         \\ \cline{2-7} 
\multicolumn{1}{|l|}{\multirow{-5}{*}{\rotatebox[origin=c]{90}{\parbox[c]{1.5cm}{TFIDF+USE}}}}     & XGBoost         & \cellcolor[HTML]{EFEFEF}\textbf{86.99} & 87.48                         & 87.19                         & 87.39                         & 94.41                         \\ \hline
\multicolumn{1}{|l|}{}                                & LSTM            & 66.74                         & 70.11                         & 67.85                         & 72.85                         & 76.19                         \\ \cline{2-7} 
\multicolumn{1}{|l|}{}                                & CNN             & 78.93                         & 81.74                         & 80.02                         & 79.89                         & 87.6                          \\ \cline{2-7} 
\multicolumn{1}{|l|}{}                                & SVM-linear      & 81.61                         & 83.79                         & 82.56                         & 82.71                         & 90.06                         \\ \cline{2-7} 
\multicolumn{1}{|l|}{}                                & Random Forest   & 80.77                         & 82.69                         & 81.45                         & 81.42                         & 88.23                         \\ \cline{2-7} 
\multicolumn{1}{|l|}{\multirow{-5}{*}{\rotatebox[origin=c]{90}{\parbox[c]{1.5cm}{TFIDF+W2V}}}}     & XGBoost         & 84.49                         & 84.49                         & 84.36                         & 84.49                         & 92.21                         \\ \hline
\multicolumn{1}{|l|}{}                                & LSTM            & 79.53                         & 81.44                         & 76.7                          & 77.54                         & 87.23                         \\ \cline{2-7} 
\multicolumn{1}{|l|}{}                                & CNN             & 86.77                         & 81.21                         & 83.37                         & 84.16                         & 92.51                         \\ \cline{2-7} 
\multicolumn{1}{|l|}{}                                & SVM-linear      & 84.24                         & 88.33                         & 86.17                         & 86.02                         & 93.36                         \\ \cline{2-7} 
\multicolumn{1}{|l|}{}                                & Random Forest   & 84.7                          & 84.63                         & 84.61                         & 84.97                         & 92.77                         \\ \cline{2-7} 
\multicolumn{1}{|l|}{\multirow{-5}{*}{\rotatebox[origin=c]{90}{\parbox[c]{1.5cm}{USE+W2V}}}}       & XGBoost         & 84.78                         & 86.87                         & 85.74                         & 85.78                         & 93.65                         \\ \hline
\multicolumn{1}{|l|}{}                                & LSTM            & 74.04                         & 91.16                         & 81.25                         & 79.24                         & 90.00                            \\ \cline{2-7} 
\multicolumn{1}{|l|}{}                                & CNN             & 85.66                         & 83.32                         & 84.27                         & 84.72                         & 92.37                         \\ \cline{2-7} 
\multicolumn{1}{|l|}{}                                & SVM-linear      & 85.21                         & 90.77                         & 87.86                         & 87.63                         & 94.15                         \\ \cline{2-7} 
\multicolumn{1}{|l|}{}                                & Random Forest   & 85.65                         & 84.84                         & 85.12                         & 85.46                         & 92.27                         \\ \cline{2-7} 
\multicolumn{1}{|c|}{\multirow{-5}{*}{\rotatebox[origin=c]{90}{\parbox[l]{1.6cm}{\centering TFIDF+USE\\+W2V}}}} & XGBoost         & 85.07                         & 87.99                         & 86.43                         & 86.42                         & 93.89                         \\ \hline
\multicolumn{2}{|l|}{RoBerta}                                           & 86.64                         & \cellcolor[HTML]{EFEFEF}\textbf{95.31} & \cellcolor[HTML]{EFEFEF}\textbf{90.77 }& \cellcolor[HTML]{EFEFEF}\textbf{90.31} & \cellcolor[HTML]{EFEFEF}\textbf{94.93} \\ \hline
\end{tabular}
\end{table}

As discussed in Section \ref{sec:models}, we developed 6 ML and 3 DL classifiers. Given we have limited space, we only present the results of the top three best-performing ML classifiers (SVM-linear, Random Forest, and XGBoost) and 3 DL classifiers (CNN, LSTM, RoBERTa) in Table \ref{tbl:MLResults}. The evaluation of other ML classifiers (Decision Tree, AdaBoost, and Logistics Regression) are available in our replication package \cite{replication}. Our experiments show that ML and DL classifiers achieve, on average, above 80\% in all metrics: precision = 80.69\%, recall = 83.54\%, F1-score = 81.66\%, accuracy = 81.68, and AUC = 88.45\%. Among ML and DL classifiers that use feature extraction methods, those that use the combination of different feature extraction methods lead to better results compared to applying them individually. The improvements range from 3.5\% to 4.5\%. Overall, results show that RoBERTa results in the best score for all the measures (precision = 86.64\%, recall = 95.31\%, F1-score = 90.77\%, accuracy = 90.31\%, and AUC=94.93\%), except for precision, without a need for a pre-processing feature extraction step. XGBoost with a combination of TFIDF and W2V feature extraction methods archives a slightly better result in the precision metric (i.e., 86.99\%) than RoBERTa. Among the other methods, SVM-linear together with a combination of TFIDF, USE, and W2V feature extraction methods, has comparably high accuracy and F1-score (accuracy = 87.63\%, F1-score = 87.86\%), followed by XGBoost with a combination of TFIDF + USE (accuracy = 87.07\%, F1-score = 87.23\%).

\subsubsection{Comparison with Baseline}  Our analysis shows that currently, no state-of-the-art approaches exist to identify gender reviews. Hence, similar to other studies (e.g., \cite {abdalkareem2020machine,9796360,alomar2021finding}), we compared our best-performing classifier (RoBERTa) with a random classifier. As we have two classes (i.e., gender reviews and non-gender reviews), the recall of the random classifier would be 0.5. The precision of the random classifier is calculated by dividing the number of gender reviews by the total number of reviews in our dataset (i.e., $precision= \frac{420}{1440}$ = 0.29). The F1-score of the random classifier is calculated at 0.36. RoBERTa outperforms the random classifier with an improvement of $2.5x = \frac{90.77}{36}$ in the F1-score.



\section{Categories of Gender Discussions (RQ2)}\label{sec:categoriegender}
\subsection{Approach}\label{sec:approachRQ2} 
To understand the types of gender discussions in reviews, we performed the open coding procedure \cite{glaser2017discovery} on 388 out of 620 gender reviews (i.e., reviews including gender discussions) collected in Section \ref{sec:dataset}. These 388 gender reviews belonged to 41 apps from 16 categories and were randomly selected as a representative sample by applying the power statistics with a 99\% confidence level and a 4\% margin of error on 620 gender reviews. In the first step, three analysts (three authors) conducted the open coding procedure to independently analyse randomly selected 100 gender reviews out of 388 gender reviews. At the end of this step, the analysts arranged multiple meetings to talk about and calibrate the generated codes, concepts, and categories and resolve any disagreements. 

In the second step, the remaining 238 reviews were divided between two of the analysts (150 reviews for the first analyst and 138 reviews for the second analyst). While the two analysts used the codes, concepts, and categories generated in the first step as a basis to guide analysing their assigned reviews, they were free to find and record new codes, concepts, and categories. Next, several meetings were organised between the three analysts to discuss the results and resolve any disagreements and conflicts using the negotiated agreement approach \cite{campbell2013coding}. In the end, all the analysts agreed upon on finalised version of gendered codes, concepts, and categories.
\subsection{Results}\label{sec:resultsRQ2}
\begin{table}[]
\caption{Categories and sub-categories of gender discussions in app reviews}
\label{tbl:categories}
\footnotesize
\begin{tabular}{m{2cm}m{6cm}}
\hline
\rowcolor[HTML]{C0C0C0} 
\textbf{Categories}                                              & \textbf{Concepts}                                                           \\ \hline
                                                        & Content management                                                          \\
                                                        & Voice settings                                                            \\
                                                        & Health and fitness                                                          \\
                                                        & Translation                                                                 \\
                                                        & User account setting                                                        \\
                                                        &  Face recognition                                                            \\
\multirow{-7}{*}{\footnotesize{\textbf{App Features}}}                & Privacy and safety concerns                                           \\ \hline
                                                        & Font, look and color                                                   \\
\multirow{-2}{*}{\footnotesize{\textbf{Appearance}}}                    & Emojis                                       \\ \hline
                                                        & Offering gender-specific, inclusive, or non-biased content                 \\
                                                        & Unbalanced amount of content on specific gender types (too much or little) \\
\multirow{-5}{*}{\rotatebox[origin=c]{0}{\parbox[c]{2cm}{\footnotesize{\textbf{Content}}}}}                      & Gender violent, sexist or inappropriate content                            \\ \hline
                                                        & Gender-biased  or sexist policy and censorship                              \\
                                                        & Neutral (fair) gender-related policy and censorship                         \\
\multirow{-3}{*}{\rotatebox[origin=c]{0}{\parbox[c]{2cm}{\footnotesize{\textbf{Company Policy \& Censorship}}}}} & Frustrations with promoting specific gender types                           \\ \hline
                                                        & Disliking gender-related Ads                                                \\
                                                        & Receiving irrelevant Ads to the gender of user                              \\
\multirow{-3}{*}{\footnotesize{\textbf{Advertisement}}}                & Unbalanced amount of gender-related Ads                                     \\ \hline
                                                        & Gender-biased or violent community                                          \\
\multirow{-2}{*}{\footnotesize{\textbf{Community}}}                    & Good or inclusive community                                                 \\ \hline
\end{tabular}
\end{table}
Our qualitative analysis of a sample of gender reviews resulted in identifying 6 high-level categories of gender discussions as presented in Table \ref{tbl:categories} and explained below.

\subsubsection{App Features} This category includes the reviews that are providing feedback on features of the app. They include raising concerns about bugs or requesting new features and sharing preferences. Our analysis revealed a significant number of gender discussions were coded under this category, including issues about content management, voice and facial recognition features, user account settings, text translation, health and fitness features, and privacy and safety issues (See Table \ref{tbl:categories}). 

We observed several comments indicating the need for more user control on searching, sorting, and receiving suggestions on gender-specific content. Sorting music tracks based on the gender of artists or labelling movies that have a specific gender audience are some examples of requests in this area. For instance, it was mentioned:

\faThumbsDown{} \textit{Please give your users greater control of filtering what they want to see and cut out what offends them. The best feature you could easily offer is searching based on religious, racial, and gender preferences. I will personally give you subscription fees if you do this.”}

Several users raised their preferences in choosing between male vs. female voices when interacting with bots. These requests were raised about voice capabilities embedded in a variety of tools such as translators, navigation tools and virtual assistants. Some apps did not offer the option of changing the default voice. Also, sometimes even the option of switching the voice existed, but it was not perceived as easy to use. However, we also observed that users appreciated it when the app offers an intuitive and easy solution to switch between voices of different genders. For example,


\faThumbsUp{} \textit{“One day I told [app name] to talk in male voice not in female voice and she did it so clearly and perfectly!! and I loved that so much... I want to say thank you for this.”} 

Besides the voice preference, we observed bugs reported that virtual assistant technologies are not as accurate or sensitive to the female voice as they are to the male voice, which might be due to biases in training data. For example:

\faThumbsDown{} \textit{“Assistant regularly can't recognise my girlfriend to add things to the shopping list (algorithmic bias towards male voices?) but then knows who she is when asked.”}

Various users shared frustrations and issues about account settings and the sign-up process of apps. We observed several comments raising requests to have more inclusive options on the choice of gender, such as:

\faThumbsDown{} \textit{“This game is pretty fun [...], what I'm a little disappointed about is choosing the gender when you sign up. There's only a Boy and Girl option but I think [it] should add custom gender too.”} 

Also, some users stated that they preferred not to share their gender with the app, for example:

\faThumbsDown{} \textit{“I find it a bit unreal that you have to choose between female and male when I do not want to specify myself with a specific gender, I rather not say it, so I would appreciate [it] if this could be a future feature that can be added...”} 

Furthermore, our data analysis revealed issues that users face with text translation features of apps without proper consideration of gender. While in some languages gender is a crucial part of grammar (e.g., influences format/ choice of verbs, and nouns), the users found it inaccurate or less useful when the translation feature does not provide any clue of gender. For instance, one of the users mentioned: 

\faThumbsDown{} \textit{“For gender(ed) languages like German, it is required to know the gender with the word, but the app doesn't display them making usefulness limited.”} 

In the context of health tracking and fitness apps, we observed some users raised concerns about lack of inclusiveness and missing non-binary users when it comes to tips, health tracking and reminders. We found that users appreciated the gender-specific tips offered by these apps, though some users requested more specialized features.

\faThumbsDown{} \textit{“It could be improved if the feminine option comes with during pregnancy and after pregnancy tips and so on....”} 

Lastly, we observed some gender reviews that raised concerns about the privacy or safety of the users. Choosing the gender of the driver or delivery person for safety reasons, and hiding gender on the app to avoid unwelcome messages from other users are among the requests in this area. 

\faThumbsDown{} \textit{“Sometimes [they] aren't reliable and as a female who used to go out at night [...] often I wish we had the option to choose the gender of our driver.”} 

\subsubsection{Appearance} This category includes those reviews that are related to the appearance of the app. Asking for a more inclusive set of emojis with options for all genders was one of the most common requests. We also observed issues raised about the auto-correction feature of apps that changes the gender of emojis chosen by users. For example:

\faThumbsDown{} \textit{“As a non-binary person, I preferred being able to use the female reaction emojis (shrug, hand up, etc.) But recently messenger has been auto-correcting these emojis into the male version based on what my profile says. If you can revert this, I'll update it to 5 stars.”} 

Some users did not like the user interface design, choice of fonts, colour and themes in the app and needed the option to choose a more masculine or feminine feel and look.

\faThumbsDown{} \textit{“You can't customize and make it different, the new colours are terrible pastel colours and I am a dude and that's really girly! Deleting the app.”} 

However, we also found comments that users appreciated when apps provided options that intuitively consider the gender of users in the design of the user interface. For example:

\faThumbsUp{} \textit{“Love the personalized stickers. It uses a picture and some AI to generate them, and I was pleasantly surprised to find that it actually gendered me correctly.”} 

\subsubsection{Content} This category comprises reviews that do not directly refer to a feature of an app but share gender-related comments about the content of the app, such as music tracks, movies, cartoons, games, and books. The comments in this category show that the reaction of users to the app content varies in a broad spectrum. Several users stated their satisfaction with the content that the app offers for different genders. This could vary from liking feminist books, and masculine/ girlish games, to movies with non-binary characters.
%
Some of the users especially liked the inclusiveness of the app, and the possibility to find content suitable for any gender. For example, a user stated: 

\faThumbsUp{} \textit{“I really like this app. It doesn't 'age' me and 'gender' me like that other app [...]. I like what I like and [it] enables me to find anything I'm looking for, quickly.”} 

Our analysis, however, revealed significant review comments indicating disappointment of users from the app contents. While some users complained about having too much content for a specific gender type, some others stated limitations in this regard. For instance, it was mentioned:

\faThumbsDown{} \textit{“ ... [it] needs to add the newest shows and old shows plus more lgbtq+ shows.”} 

Furthermore, we also found some comments that users shared their frustrations with the content of the apps due to being sexist, gender violent, inappropriate for children, or being against their beliefs and value systems. 




\subsubsection{Company Policy and Censorship} This category includes those reviews that are raising concerns about gender-related policies of companies or app owners and their approach towards censorship. Like the “Content” category, we observed a variety of comments being in favour or against the gender-related policies of the company. Some of the users were concerned about companies promoting specific genders. Some others shared frustrations about companies making biased decisions when it comes to censoring content. For example, one of the users posted:

\faThumbsDown{} \textit{“For some reason it takes your videos down a lot when you’re a person of colour, neuro-divergent, or LGBT.”} 



We also observed comments that users found an app to be fair, neutral, and not biased in censoring or promoting gender-specific content, such as: 

\faThumbsUp{} \textit{“Excellent on all levels and most importantly Gender, Race, and equality and each person's location!”} 

\subsubsection{Advertisement} Under this category, we have reviews that are sharing feedback about advertisements that users receive via apps. Most of the comments in this category reflected the frustration of users with the ads they receive. Some users felt overwhelmed by receiving too many ads about gender-specific goods or events. Some users specifically disliked the gender-related Ads and felt uncomfortable receiving them. For instance, one of the users stated:

\faThumbsDown{} \textit{“The ads were fine until recently and they started spewing propaganda onto everything. Race, religion, gender. I have since uninstalled [it] and won't be back. [...] I should be able to make my own thoughts without you being a sellout to mainstream media, and trying to influence me or my peers.”} 

Some others complained about receiving ads, that are not relevant to their gender. For example: 

\faThumbsDown{} \textit{“So many ads that aren't even aimed at me… it's almost like you're trying to ‘force’ me into your way of thinking […] you're ruining this app for straight biological men.” } 

\subsubsection{Community} The reviews under this category are sharing feedback about the community of the app and their gender-related behaviour. We observed some comments complaining that communities on apps (e.g., social media, gaming groups) are violent or share hateful comments against specific genders. Some users expected better moderation from the app owners to remove these contents regularly, ban the accounts, or facilitate reporting such behaviours. On the other hand, we also observed positive comments about inclusiveness, diversity and how the app has enabled users to connect with their preferred community. For example, a user mentioned:

\faThumbsUp{} \textit{ “... introduced me to some of the best people I’ve met. […] helped me learn about transgender from people […].”} 


\section{Discussion and Implications}\label{sec:discussion}
This section presents our reflections on the findings and recommendations and implications for research and practice.
\\
\\
\noindent \textbf{\textit{User empowerment}.} Our study reveals that many users are upset about not having any control and freedom over gender-related features, advertisements they receive from apps, and the content suggested by apps. While several confounding factors (e.g., personal data leakage) can demotivate users to accept and use mobile apps \cite{stocchi2019drivers,harris2016identifying}, we found gender-related concerns could lead to the situation that some users decide not to use the app anymore. We argue that the role of gender-related concerns in user satisfaction can be attributed to the fact that gender perspective is a complex phenomenon, which impacts and is influenced by different characteristics of users such as their religion, language, background, and culture \cite{nunes2023gire}. Given that the diverse users of an app have varied gender perspectives, they want to have more freedom and control over the app's gender-related features, advertisements, and content. 

\textbf{\underline{Implications}}. Our research highlights that mechanisms should be developed by app developers to empower users to apply their gender-related preferences and expectations in apps in a meaningful way. It is also equally important to have gender diversity in the software development teams to make sure such gender-related preferences and expectations are being understood. Further research is also needed to guide the design and evaluation of such mechanisms.     
\\
\\
\noindent \textbf{\textit{Need more diverse and inclusive user-based research}.}
Our study indicates that some users express gender-related concerns about policies adopted by the app owners, algorithms employed to recommend the content (e.g., movies), and receiving advertisements that are not aligned with their gender. The lack of knowledge about users' perspectives on gender among app developers can be one of the main causes behind the identified gender issues. Nunes et al. \cite{nunes2023gire} highlight that the software engineering community has limited approaches, that can enable practitioners to collect and understand users' perceptions of gender at the early stage of software development. Further to this, as discussed earlier, gender perspectives can vary by religion, culture, age, etc. All this makes it difficult for software/app developers to understand the impacts of gender on apps and propose solutions to address gender-related concerns. 

\textbf{\underline{Implications}}. While our research serves as the foundation for understanding the types of gender discussions expressed by app users in mobile app reviews, more diverse and inclusive user-based research is needed to gain deep insights into the perceptions of users on gender. As an example, future research can compare gender perspectives and related concerns in different classes (e.g., different geographical locations, cultural backgrounds, and ages). The outcomes of such inclusive and diverse research hope to shed light on the impacts of different policies on users, indicate which ones are the best, and help developers adopt better solutions for apps' advertisements and content. We also argue that one of the challenges of developing inclusive mobile apps is providing innovative approaches for continuously collecting and monitoring users' perspectives on gender-related matters. As the apps expand in different locations and their user base scales, it is important that app owners invest in solutions to systematically observe the demography of their users and collect their feedback. In addition, we highlight that it is equally important to finding solutions for meaningfully analyzing users' feedback \cite{wang2022demystifying,martens2019towards}. Given the reviews are given based on personal interpretations and experiences of users, there is a need for more research to differentiate between constructive feedback from hateful comments, maintain the balance of different perspectives and translate the reviews into reasonable points to be addressed by app owners. To this end, a multidisciplinary research and development team, including those with different expertise in psychology, law, and sociology, is needed for meaningful analysis of user feedback (e.g., from a gender perspective).
\\
\\
\noindent \textbf{\textit{Towards automated summarisation of gender discussions in software artifacts}.}
To the best of our knowledge, our study is the first to investigate gender discussions (e.g., discussion about problems that an app is making for its user from a gender perspective) in app reviews. Given that there was no reliable and well-known list of gender-related keywords in software engineering, we used the KeyBERT method combined with manual analysis to build a unique list of gender-related keywords for our study. Further to this, while the developed binary ML/DL classifiers can automatically distinguish gender reviews from non-gender reviews with high accuracy (\textbf{RQ1}), we decided to use a manual approach to categorise the types of gender discussions in gender reviews (\textbf{RQ2}). This decision limited the size of our dataset, as the exponentially increasing size of datasets makes manual methods time-consuming and difficult. Moreover, we only focused on gender discussions in a single data source, i.e., app reviews. Gender discussions can occur at any software/app development stage and can be expressed in different resources/artefacts, such as code, issues, code comments, developer discussions, etc. Each software artifact may disclose different types of gender discussions. 

\textbf{\underline{Implications}}. We argue that summarisation approaches should be developed to automatically summarise gender discussions at scale from several sources (i.e., code, issues, code comments, developer discussions in pull requests, code commits, and app reviews). Similar to recent works on summarising privacy concerns \cite{nema2022analyzing,ebrahimi2022unsupervised}, summarisation approaches for gender can be built on top of our developed binary classifies and use the constructed gender-related keyword set. Software/app organizations will be able to use such an integrated summarisation approach to summarise gender discussions from distinct sources instead of multiple gender discussions identifiers and summarisation approaches. Hence, such summarisation approaches would be simpler, more scalable, and easier to use.
\\
\\
\noindent \textbf{\textit{Study how the types of gender discussions appear in different categories and apps.}} This study investigated 388 gender reviews to identify 6 types of gender discussions in reviews. These 388 gender reviews come from 41 apps that belong to 16 categories (e.g., communication, finance, social, etc,). We studied gender discussions in reviews \textit{independently} of app categories, while the number of gender reviews across app categories was highly variable. More specifically, 217 (56\%) out of the 388 gender reviews are from only 4 categories: \textit{Communication}, \textit{Tools}, \textit{Social}, and \textit{Entertainment}. However, we argue that a type of gender discussion might be more prominent in some categories while appearing in other categories rarely \cite{noei2019too}. For example, gender discussions around the “Community” topic may less be reported in apps in the “\textit{Finance}” category. Moreover, from the perspective of users and/or app owners, the importance of each type of gender discussion may vary in each app category. For example, the users of apps like YouTube, which is classified in the “\textit{Video Players \& Editors}” category, might have more gender-related concerns about the content of the app or advertisements they receive than the community of the app.

\textbf{\underline{Implications}}. More research is needed to investigate the emergence, frequency, and importance of types of gender discussions in different app categories. The results of such studies can inform researchers and app owners about the prominent and important aspects of gender discussions in relation to app business domains and enable them to allocate and prioritise times and efforts to understand and address gender-related concerns and requirements.


\section{Threats to Validity}\label{sec:Threats}
Here, we discuss the threats to the validity of this study and the adopted strategies (if any) to mitigate these threats \cite{wohlin2012experimentation}.

\textbf{Internal Validity}. The process of building a dataset of gender reviews may have introduced some threats. First, we decided to only investigate the top 70 Android popular apps in the Google Play Store according to the 2018 App Annie Report \cite{annie2020app} and collect 7 million most relevant reviews from these apps (Section \ref{sec:dataset}). Other researchers also used this report and explored the most popular apps listed in the report (e.g., \cite{chen2021should}). While the 70 apps are from a wide range of categories, we confirm that we might have missed Android apps and user reviews that were important from the gender perspective. Our approach to finding potential gender reviews in Section \ref{sec:dataset} may be limited to the accuracy of the initially chosen keywords and the KeyBERT method. While KeyBERT is a simple and widely used method to extract keywords from a document, we mitigated the concern about the accuracy of KeyBERT by three strategies: (a) applying KeyBERT on other relevant datasets apart from app reviews, (b) manually analysing the extracted keywords, and (c) manually analysing 10 reviews corresponding to every extracted uni-gram keyword. Overall, we believe that our keyword set is comprehensive enough and the final dataset includes the vast majority of possible gender-related discussions. Despite this, we acknowledge we might have missed some cases of gender-related discussions.

The qualitative analysis of app reviews to label them into gender reviews or non-gender reviews might be subjective. We reduced this threat by (a) involving three analysts (coders) who had a solid knowledge of human-centric software engineering and asking them to independently label each review and (b) using the negotiated method as a well-established method in the qualitative research to solve disagreements among coders. Furthermore, when the three analysts were unsure about the type of review, they labelled it as a non-gender review. The analysis of gender reviews to identify the types of gender discussions is subjected to the bias of analysts (\textbf{RQ2}). This threat was mitigated by (a) involving three analysts in the analysis process (i.e., investigator triangulation \cite{carter1969use}), (b) conducting this process in three iterations to avoid fatigue and develop a common understanding among the analysts, and (c) ensuring all the analysts agree on the final types of gender discussions.

\textbf{Construct Validity}. Similar to other studies, we used a limited number of ML/DL models (i.e., 9 models) \cite{peters2017text}. Except for RoBERTa, we also only used three feature extraction techniques and their combinations in the remaining ML/DL models. While these feature extraction techniques and ML/DL models are widely used in research and practice communities, we acknowledge that the use of other models and feature extraction techniques could yield different results. Further to this, while we fine-tuned the parameters of the ML/DL models, changing the parameters may lead to different results. We also acknowledge that the precision of the best-performing classifier should be improved to be used in real-world settings. Among many metrics that could be used to evaluate the performance of ML/DL models, we used five commonly used ones: precision, recall, F1-Score, accuracy, and AUD \cite{yang2022survey}. Finally, we only collected our data from one source (i.e., app reviews). Besides mining reviews, other research methods such as surveying app users could be used to triangulate the findings of RQ2 \cite{carter1969use}.

\textbf{External Validity}. Our research in this paper is based on the top 70 popular Android apps in the Google Play Store listed in the 2018 App Annie Report \cite{annie2020app}. Furthermore, we only investigated a subset of reviews in these 70 apps (i.e., 100K most relevant reviews in each). We also did not include any iOS apps from the Apple App Store and their corresponding reviews in our study. Finally, only a small set of gender reviews (388 reviews) was manually analysed to answer RQ2. All of this indicates that the types of gender discussions identified in RQ2 may not be comprehensive and generalisable to all app reviews and apps in the Apple App Store and Google Play Store and to any other type of software systems. However, we argue that this threat was partially reduced as the Android apps used in this study come from a variety of domains/categories and receive a very large number of reviews, which can to some extent indicate their popularity and representativeness \cite{gao2022understanding}. 

\textbf{Reliability}. Other researchers may want to replicate our study but there is a threat that they obtain different results. Hence, we created a replication package and released it online \cite{replication} for any further research. The replication package includes the keyword set developed in Section \ref{sec:dataset}, the dataset to evaluate the performance of ML/DL models and identify the types of gender discussions, and the source code of ML/DL models.

\section{Conclusion and Future Work}\label{sec:conclusion}
In this work, we studied gender discussions from the user perspective in mobile app reviews in the Google Play Store to provide a basis to develop gender-inclusive apps. We first developed six ML and three DL classifiers that automatically detect app reviews that contain discussions around gender (i.e., gender reviews). The ML and DL classifiers were evaluated on a manually constructed dataset of 1,440 app reviews, consisting of 620 gender reviews and 820 non-gender reviews. We found among ML and DL classifiers, RoBERTa as a DL classifier does this task best, with a precision of 86.64\%, recall of 95.31\%, F1-score of 90.77\%, accuracy of 90.31\%, and AUC of 94.93\%. Additionally, we qualitatively analysed 388 gender reviews to develop a deep insight into the types of gender discussions in app reviews. We found that users of mobile apps discuss their gender concerns around six topics in app reviews: \textit{App Features}, \textit{Appearance}, \textit{Content}, \textit{Company Policy and Censorship}, \textit{Advertisement}, and \textit{Community}.
 
In the future, we plan to explore and develop practices that can be employed to consider and include gender in mobile app development. We also aim to explore gender discussions in other software repositories. Finally, we will look into the correlation between the six types of gender discussions and other factors such as user satisfaction, the categories of apps, and the gender of reviewers, through sentiment analysis.

 


\bibliographystyle{IEEEtran}
\bibliography{mainref.bib}
\end{document}